\documentclass[aps,prb,reprint,groupedaddress,showpacs,amsmath,amssymb]{revtex4-1}
\usepackage{graphicx}
\usepackage{dcolumn}
\usepackage{bm}
\usepackage{float}

\begin{document}


\title{ Temperature-driven band inversion in Pb$_{0.77}$Sn$_{0.23}$Se: Optical and Hall-effect studies }


\author{Naween Anand}
\affiliation{Department of Physics, University of Florida, Gainesville, FL 32611-8440, USA}
\author{Zhiguo Chen}
\affiliation{NHMFL Florida State University, Tallahassee , FL 32310-3706, USA}
\author{Zhiqiang Li}
\affiliation{NHMFL Florida State University, Tallahassee , FL 32310-3706, USA}
\author{Sanal Buvaev}
\affiliation{Department of Physics, University of Florida, Gainesville, FL 32611-8440, USA}
\author{Kamal Choudhary}
\affiliation{Department of Materials Science and Engineering, University of Florida, Gainesville, FL 32611-8440, USA}
\author{S. B. Sinnott}
\affiliation{Department of Materials Science and Engineering, University of Florida, Gainesville, FL 32611-8440, USA}
\author{Genda Gu}
\affiliation{Condensed Matter Physics and Materials Science Department, Brookhaven National Laboratory, Upton, NY 11973-5000, USA}

\author{C. Martin}
\affiliation{Department of Physics, Ramapo College, Mahwah, NJ 07430, USA}
\author{A. F. Hebard}
\affiliation{Department of Physics, University of Florida, Gainesville, FL 32611-8440, USA}
\author{D. B. Tanner}
\affiliation{Department of Physics, University of Florida, Gainesville, FL 32611-8440, USA}


\date{\today}

\begin{abstract}
Optical and Hall-effect measurements have been performed on single crystals
of Pb$_{0.77}$Sn$_{0.23}$Se, a IV-VI mixed chalcogenide. The temperature
dependent (10--300~K) reflectance was measured over 40--7000~cm$^{-1}$
(5--870~meV) with an extension to 15,500~cm$^{-1}$ (1.92~eV) at room temperature.
The reflectance was fit to the  Drude-Lorentz model using a single Drude component
and several Lorentz oscillators. The optical  properties at the measured temperatures
were estimated via Kramers-Kronig analysis as well as by the Drude-Lorentz fit.
The carriers were p-type with the carrier density determined by Hall measurements.
A signature of valence intraband transition is found in the low-energy optical spectra.
It is found that the valence-conduction band transition energy as well as the free carrier
effective mass reach minimum values at 100~K, suggesting temperature-driven band inversion
in the material. Density function theory calculation for the electronic band structure also make similar predictions.
\end{abstract}
\maketitle

\section{Introduction}

Class IV-VI narrow-gap compound semiconductors have been of great interest for many
decades not only for their scientific interest but also for their use in novel
technological instrumentation such as infrared optoelectronics and thermoelectric devices.\cite{Nimtz}
Crystalizing in the rock salt structure, materials such as PbTe, PbSe, SnTe, and their mixed alloys
have been found to have high dielectric constants and quite unusual infrared and electronic properties.\cite{Otfried}
PbSe and SnSe, the two parent compounds of Pb$_{1-x}$Sn$_{x}$Se alloy system, have what is commonly known as band inversion.\cite{Dimmock}
For  $x$ = 0, $i.e.$, PbSe has the conduction band in L${}_{6}^{+}$ symmetry and the valence band symmetry is denoted by L${}_{6}^{-}$.
 As one increases $x$ in the alloy system, Pb$_{1-x}$Sn$_{x}$Se, the band gap initially reduces and then closes, with a linear
 dispersion around the Fermi level. Subsequent increase in $x$  reopens the gap but now the valence band symmetry gets inverted
 to L${}_{6}^{+}$, whereas symmetry of the conduction band changes to L${}_{6}^{-}$, identical to SnSe, $i.e.$, $x$ = 1.

Recent theoretical and experimental interest in these alloys has occurred because they have been suggested to represent
 a new, non-trivial topological phase called a topological crystalline insulator\cite{Fu,Hsieh}(TCIs).
 As in most topological insulators, observations are complicated because the crystals are usually found to be either
  p-doped or n-doped due to non-stoichiometry so that surface-state features get overshadowed by an overwhelming bulk-carrier contribution.
 This is the case in the data reported here. Nevertheless, the temperature dependence of the free-carrier effective mass and of the
  valence-to-conduction band absorption edge could be a suitable point of inquiry to explore the band inversion and temperature-driven
  phase transition in such materials. So, apart from determining the optical and transport properties, this study also attempts to
  investigate such a possibility in a Pb$_{0.77}$Sn$_{0.23}$Se single crystal.

Studies of PbSe have reported the rock salt crystal structure at ambient temperature and pressure with a lattice parameter
 of $a = 6.13~${\AA} and a direct minimum energy band gap of around 0.28~eV at the L point in the Brillouin zone.\cite{Delin,Otfried,Harman}
 In contrast, SnSe has been stabilized as an orthorhombic crystal with layered symmetry and an indirect minimum energy band gap of 0.9~eV.\cite{Otfried,Albers}
  On the basis of X-ray diffraction studies of annealed powders, it has been established\cite{Littlewood,Dixon}
   that the mixed alloy Pb$_{1-x}$Sn${}_{x}$Se stabilizes in the rock salt structure for $0 \leq x \leq 0.43$ and that the minimum band gap
    remains at the L point. Infrared absorption measurements\cite{Strauss1967}  find that the gap is a complex function of
    temperature and stoichiometric ratio $x$. An ARPES study\cite{Dziawa} of Pb$_{0.77}$Sn${}_{0.23}$Se finds that the band gap
    is temperature-dependent, with a minimum around 100~K. The bulk band gap reopens at the L point with an inverted symmetry
     of the valence band L${}_{6}^{+}$ and the conduction band L${}_{6}^{-}$ as the temperature is further increased.\cite{Dziawa}

\section{EXPERIMENTAL PROCEDURES}

Compositional and structural defects play significant role in determination of chemical and electronic properties of
lead chalcogens. Local compositional disorder significantly affects the band gap formation for the alloy compositions.
In addition, short-range disorder breaks the degeneracy near the band edge\cite{Xing} because PbSe and SnSe represent
two different crystal structures. Therefore, controlled crystal growth is an essential requirement.

A vapor-phase growth technique was employed to prepare single crystals of Pb${}_{0.77}$Sn${}_{0.23}$Se. The crystals
had the rock salt structure with lattice parameter $a = 6.07~${\AA}. To ensure good quality samples, quality-control
conditions were used, including super-saturation of gaseous phase and a seed crystal with the proper structure and orientation.
 Lead chalcogen crystals are opaque and have a metallic luster. They are brittle and easily cleave along the (100) plane.
 Chemical binding in these systems has both ionic and covalent components. A single crystal of size $4\times4\times2$ mm${}^{3}$
 with smooth (100) crystal plane as the exposed surface was selected for optical measurements.

Temperature dependent (10--300~K) reflectance measurement were conducted using a Bruker 113v Fourier-transform interferometer.
 A helium-cooled silicon bolometer detector was used in the 40--650 cm${}^{-1}$ spectral range and a DTGS detector was used
 from 600--7000 cm${}^{-1}$. Room temperature measurement up to 15,500 cm$^{-1}$ used a Zeiss microscope photometer.
  Because the rock salt structure of the material implies isotropic optical properties, all optical measurements were performed
  using non-polarized light at near-normal incidence on the (100) crystal plane. To achieve higher accuracy during reflectance
  measurements, a small evaporation device incorporated in the metal shroud of the cryostat was used to coat the crystal surface with gold,
   minimizing changes of experimental conditions during sample and reference single-beam spectral measurements. The gold coating was easy
   to remove since the material is easily cleaved along the (100) plane.

Optical measurements were followed by Hall-effect measurements, using a physical properties measurement system (PPMS) from Quantum Design
which allows transport measurement over 10--300~K. Magnetic fields up to 7~T were applied perpendicular to the (100) plane and ramped
in the upward and downward directions while the Hall voltage was measured transverse to current and field. The Hall-effect measurements
gave the sign (p-type) and value for the carrier density $n$ at 10~K and at room temperature. Errors in the carrier density estimate
originate from the non-uniform thickness of the sample, non-rectangular cross section, and contact size. The errors are of course the same
 at each temperature. These errors do affect the discussion in a later section, where we compare Hall measurements and the infrared plasma
 frequency $\omega_{p} = \sqrt{4\pi n e^2/m_h^*}$, where $e$ is the electronic charge and $m^*$ is the effective mass of the holes,
 Our goal in this comparison is to obtain an estimate of the effective mass. In our analysis, the carrier concentration $n$ has been linearly
 extrapolated over intermediate temperatures for the estimation of effective number of carriers $N_{\mbox{\it eff}}$ at various temperatures of
 interest. The volume expansion of the crystal with temperature is also included in this calculation, although it is not very significant.
 Because of the similarity in the crystal structure and lattice parameter, the linear thermal expansion coefficient for this material is taken
 to be the same as PbSe.\cite{Novikova}

\section{EXPERIMENTAL RESULTS}

\subsection{Hall-effect studies}

Previous Hall studies\cite{Martinez,Maier} on PbSe and SnSe reported p-type carriers in these materials. A slight increase in p-type carrier
density at room temperature compared with 70~K was observed in SnSe.\cite{Julian} The origin of the p-type extrinsic carrier in the crystal
is due primarily to the presence of ionized lattice defects associated with deviations from stoichiometry, possibly due to excess of selenium
 or vacancies occupied by acceptor impurities.\cite{Strauss1968,Anasabe,Goldberg} Our Hall measurements show  p-type carriers with a concentration
  of $n = 6\times10^{19}$~cm${}^{-3}$ at room temperature. This value increases to $n = 9\times10^{19}$~cm${}^{-3}$ at 10~K. The relatively high value
  for $n$ suggests that our sample is a degenerate semiconductor, in which the chemical potential is pushed down into the valence band due to the
   carrier density $n$ exceeding the valence band edge density of states.

\subsection{Reflectance spectra}

The temperature dependence of the reflectance of Pb${}_{0.77}$Sn${}_{0.23}$Se between 40 and 7000 cm${}^{-1}$ (5~meV--870~meV) is shown in Fig. 1.
In the far-infrared range, a high reflectance is observed, decreasing somewhat as temperature increases. The effect of an optically-active transverse
 phonon may be seen around 40--200~cm${}^{-1}$. This feature weakens as temperature increases.\cite{Habinshuti} A rather wide valence intraband
 transition centered around 650 cm${}^{-1}$ and shifting to lower frequency with increasing temperature affects the reflectance; this is evident
 at low temperatures causing a rise in the reflectance level around the central frequency.\cite{Brian} There is a plasma minimum around 930 cm${}^{-1}$
 at 10~K; this feature shifts to around 885 cm${}^{-1 }$ as temperature increases to 300~K and becomes slightly less deep. As we shall see, it is
 in good agreement with the zero crossing of the Kramers-Kronig derived real part of the dielectric function $\varepsilon_1$.

\begin{figure}[H]
\centering
\includegraphics[width=3.4 in,height=3.4 in,keepaspectratio]{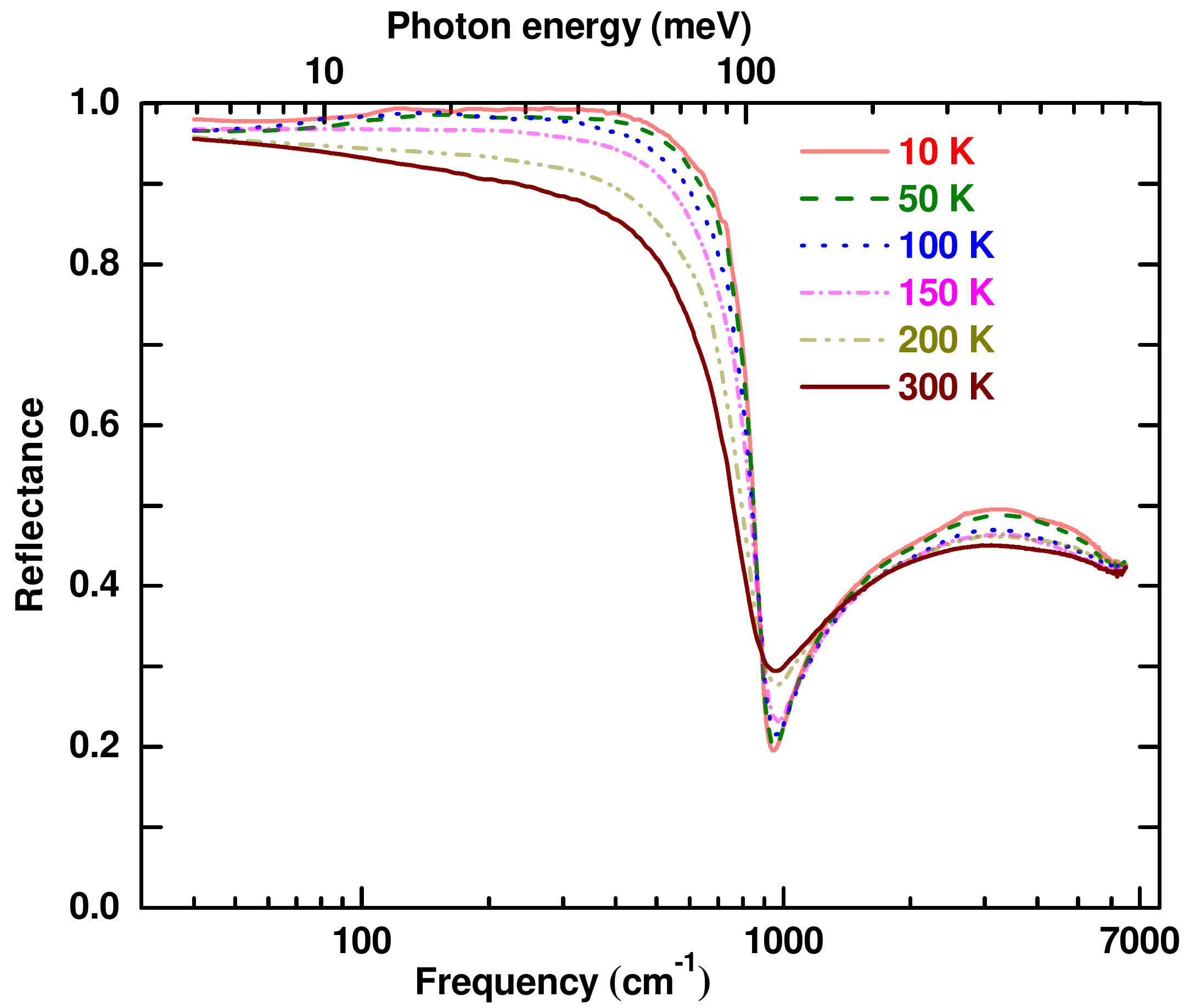}
\caption{\label{fig:Refl} (Color online) Temperature-dependent reflectance spectra of a Pb$_{0.77}$Sn$_{0.23}$Se single crystal.}
\end{figure}

In the midinfrared, we observe interband transitions from the occupied states in the valence band to the conduction band.
These transitions become weaker, leading to decreased reflectance, as the temperature is increased. As this is a p-type system,
 the chemical potential lies in the valence band; therefore, the experimentally-observed minimum absorption edge in the midinfrared
 range is an overestimate of the direct band gap at the L point of the Brillouin zone, as would be observed in an intrinsic Pb$_{0.77}$Sn$_{0.23}$Se crystal.

The top panel of Fig. 2 shows the 300~K reflectance up to 1.92 eV. There is a strong high-energy interband transition at 10,600~cm${}^{-1 }$ (1.3~eV).

\section{ANALYSIS}

\subsection{Kramers-Kronig analysis and optical conductivity}

We used the Kramers-Kronig relations to analyze the bulk reflectance $R(\omega  ) $ and then to estimate the real and
imaginary parts of the dielectric function.\cite{Wooten}  Before calculating the Kramers-Kronig integral, the low frequency
reflectance data were extrapolated to zero using the reflectance-fit parameters. (The fit is discussed later in this section.)
 Reflectance data above the highest measured frequency were extrapolated between 80,000 and $2\times10^{8}$ cm${}^{-1 }$ with the
  help of X-ray-optics scattering functions; from the scattering function   for every atomic constituent in the chemical formula
  and the volume/molecule (or the density) one may calculate the optical properties in the X-ray region.\cite{Henke} A power-law in $1/\omega$
  was used to bridge the gap between the data and the X-ray extrapolation. Finally, an $\omega^{-4}$ power law was used above
  $2\times10^{8}$~cm${}^{-1 }$. The optical properties were derived from the measured reflectance and the Kramers-Kronig-derived phase
  shift on reflection.
\begin{figure}[H]
\centering
\includegraphics[width=4.5 in, height=4.5 in, keepaspectratio]{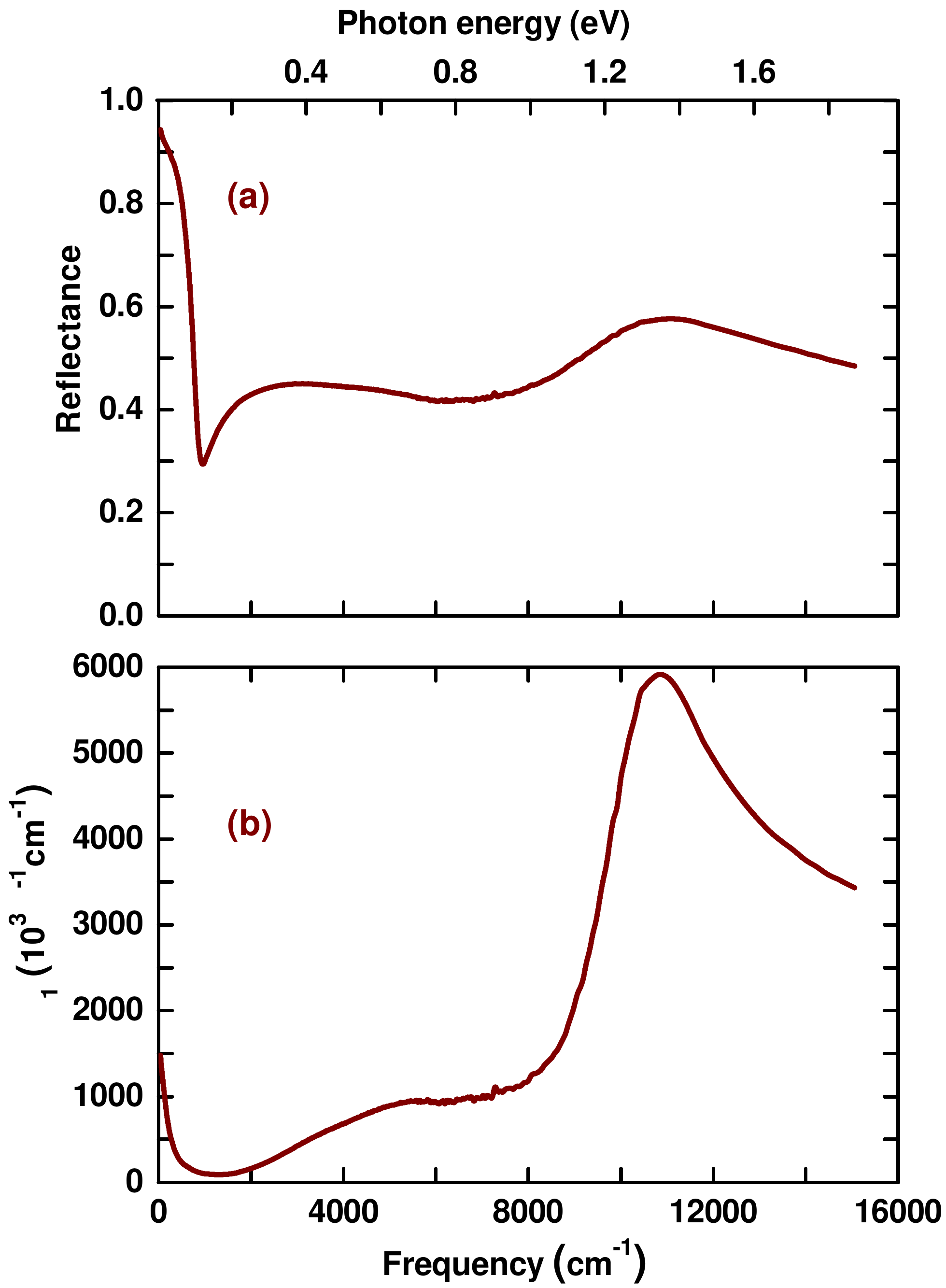}
\caption{\label{fig:300K} (Color online) (a) Reflectance spectrum from 40--15,500~cm${}^{-1}$  (5 meV--~1.92 eV) at 300K. (b) Kramers-Kronig
calculated optical conductivity from 40--15,500~cm${}^{-1}$  (5 meV--~1.92 eV) at 300K.}
\end{figure}

The lower panel of Fig.~2 shows the Kramers-Kronig-derived optical conductivity at 300~K. The narrow Drude contribution and
the contributions to the conductivity from interband transitions around 3500, 5100, and 10,600 cm$^{-1}$ are consistent with the reflectance.

The temperature dependence of the Kramers-Kronig derived real part of the optical conductivity, $\sigma_{1}(\omega)$ over 40--7000 cm${}^{-1}$
is shown in Fig. 3. The far-infrared range shows the Drude contribution to $\sigma_{1}(\omega)$. Between 10--100~K, the Drude
relaxation rate $1/\tau$ is below the measured frequency range, explaining the non-constant area under the displayed conductivity
spectrum as temperature changes. The relaxation rate increases as temperature is increased until, by 200 K, most of the Drude spectral
weight is seen in the figure.

\begin{figure}[H]
\centering
\includegraphics[width=3.5 in,height=3.5 in,keepaspectratio]{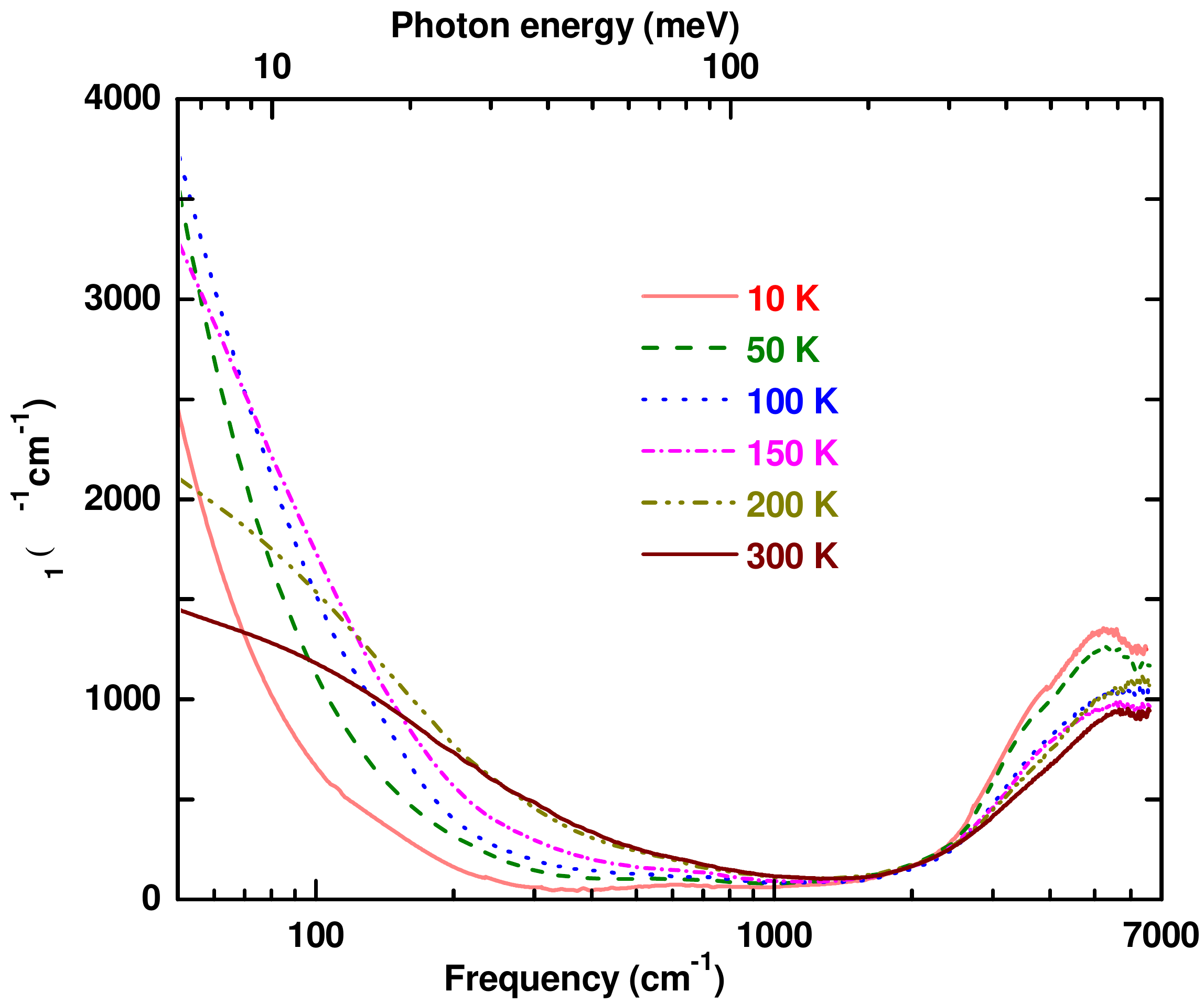}
\caption{\label{fig:conductivity} (Color online) Temperature-dependent optical conductivity of Pb${}_{0.77}$Sn${}_{0.23}$Se single crystal
obtained by Kramers-Kronig analysis.}
\end{figure}

The Drude conductivity partially overlaps with a small and dispersive contribution from the valence intraband transition at
650 cm$^{-1}$. This feature can be seen in the spectra at 10--100~K but becomes masked by the Drude spectrum as temperature increases.
 The valence intraband transition is an excitation from an occupied state below the Fermi level to an empty state (in another valence subband)
  above the Fermi level. It would only be observed if the Fermi level were to lie somewhere within the valence band.

In the midinfrared range, the optical conductivity decreases as temperature is increased. At most temperatures, two overlapping conductivity
 peaks may be discerned around 3500 and 5100 cm$^{-1}$; these features,  arising from interband transition, are also consistent with the reflectance.
 The onset of the interband transitions   is estimated by extrapolating linearly the conductivity to $\sigma_1(\omega) = 0$. This intercept is
 about 2300 cm$^{-1}$ at 10~K, then decreases to around 2200 cm$^{-1}$ at 100~K, and then increases back to 2300 cm$^{-1}$ at 300~K.
 The temperature dependence of all parameters is discussed in a following subsection.

\subsection{The real part of the dielectric function}

The Kramers-Kronig result also allows us to estimate the real part of dielectric function $\varepsilon_1$ shown in Fig. 4.
At low frequencies, $\varepsilon_1$ is negative, a defining property of a metal. The temperature dependence of the free-carrier
scattering rate is very evident in this dielectric function plot. For temperatures above 150~K, $\varepsilon_1$ becomes almost flat at low
frequencies, implying that the scattering rate is on the order of a few hundred cm$^{-1}$. In contrast, at low temperatures
$\varepsilon_1 \sim 1/\omega^{-2}$, sharply decreasing, indicating a very small scattering rate, one below the minimum measured frequency.

\begin{figure}[H]
\centering
\includegraphics[width=3.5 in,height=3.5 in,keepaspectratio]{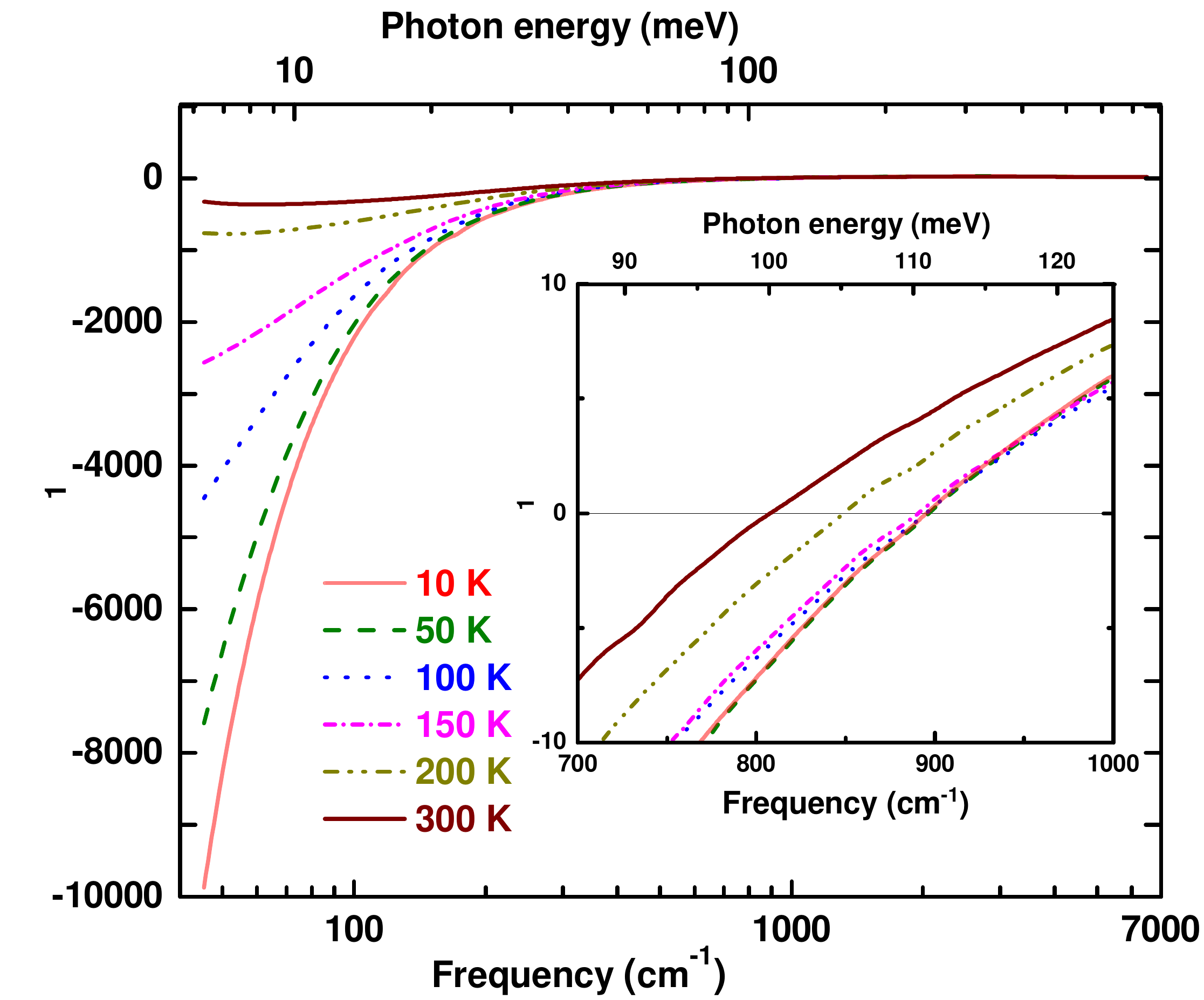}
\caption{\label{fig:epsilon} (Color online) Temperature dependence of the real part of the dielectric function of Pb${}_{0.77}$Sn${}_{0.23}$Se.
The inset shows the zero crossing of $\varepsilon_1(\omega)$, illustrating the temperature dependence of the screened plasma frequency.}
\end{figure}

The inset graph shows the zero crossing of $\varepsilon_1$ (representing the screened plasma frequency) for different temperatures.
 At 10~K the screened plasma frequency is around 900 cm$^{-1}$; this value is maintained to about 150~K, after which  it decreases to
 around 815 cm$^{-1}$ by 300~K. This Kramers-Kronig-derived screened plasma frequency is in good agreement with the reflectance plasma minimum.

\subsection{Sum rule analysis and \boldsymbol{${N_{\mbox{\it eff}}\frac{m}{m_h^{*}}}$} }

Once we have the real part of the conductivity, we can calculate the partial sum using
\begin{equation}
N_{\mbox{\it eff}}\frac{m}{m^{*}}=\frac{2mV_{c}}{\pi e^{2}}\int^{\omega}_{0}\sigma_{1}(\omega^{\prime})d\omega^{\prime},
\end{equation}
to obtain the effective number of carriers participating in optical transitions up to frequency $\omega$. The quantities in this equation are $m^*$,
 the effective mass, $m$, the free-electron mass, $e$, the electronic charge, and $V_c$, the volume taken up by one ``molecule'' of Pb${}_{0.77}$Sn${}_{0.23}$Se.

Figure 5 shows the temperature-dependent partial sum rule results. The free-carrier contribution saturates around 0.015 at 10~K, decreasing to 0.011
 at 300~K. These data suggest that either the carrier density is decreasing or the effective mass is increasing as temperature increases. The Hall
 results tell us that it is the former. We use the Hall carrier density to estimate the free carrier effective mass and its temperature dependence.
  The results are shown in Fig. 6. A uniform error is added to the data points based on the uncertainty in the $N_{\mbox{\it eff} }$ and the optical
  conductivity-dependent partial sum rule derived from Karamers-Kronig relations.

\begin{figure}[H]
\centering
\includegraphics[width=3.4 in,height=3.6 in,keepaspectratio]{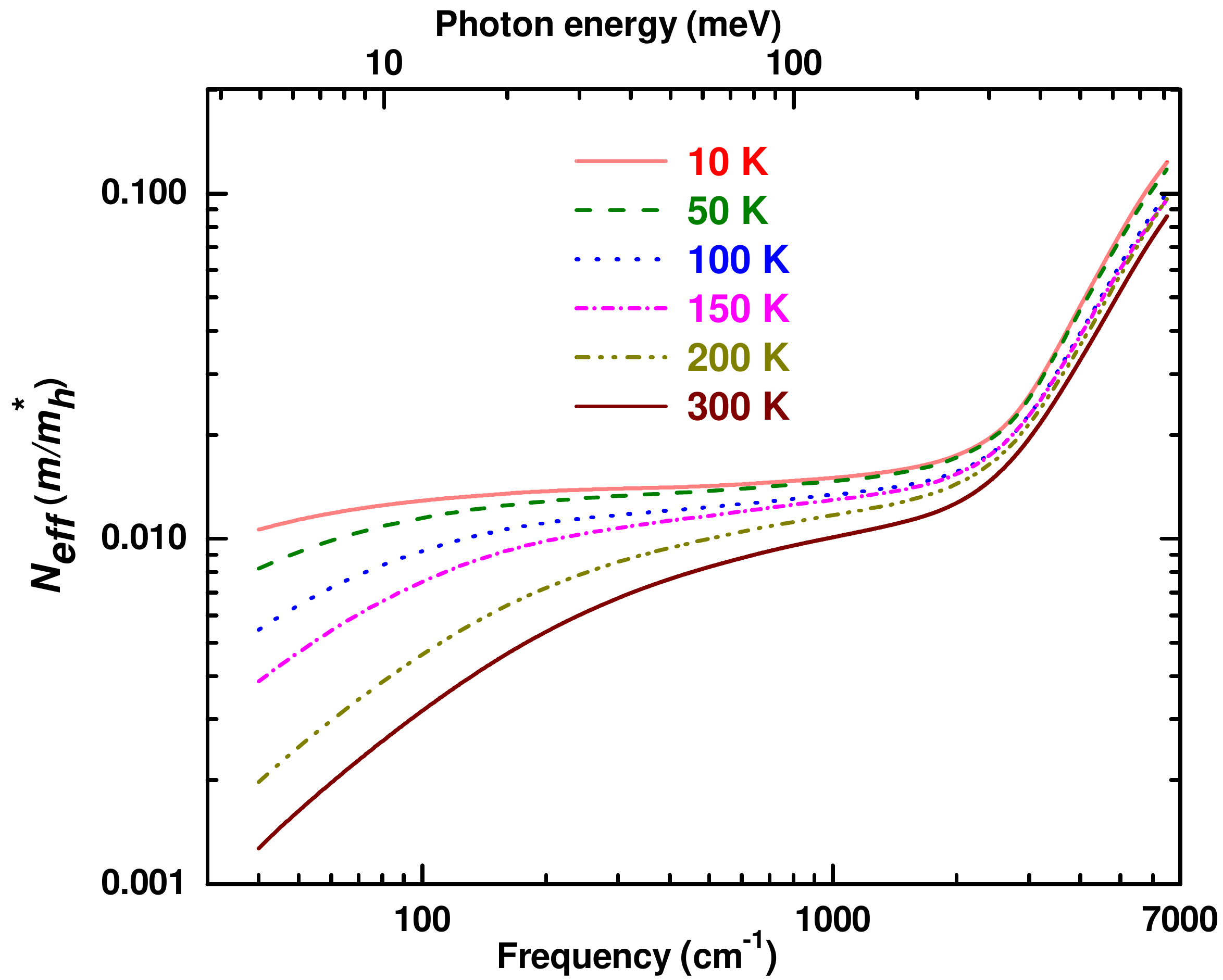}
\caption{\label{fig:Sumrule} (Color online) Temperature-dependent integrated spectral weight of Pb${}_{0.77}$Sn${}_{0.23}$Se.}
\end{figure}

The effective mass changes only slightly over the range of measurements; there is a weak minimum at 100~K where $m_{h }= 0.335 m_{e }$ and then
a rise to $m_{h }= 0.350 m_{e }$ at 300~K.  The average effective mass is $\langle m_{h }\rangle_T = 0.341 m_{e}$.

\begin{figure}[H]
\centering
\includegraphics[width=3.4 in,height=3.4 in,keepaspectratio]{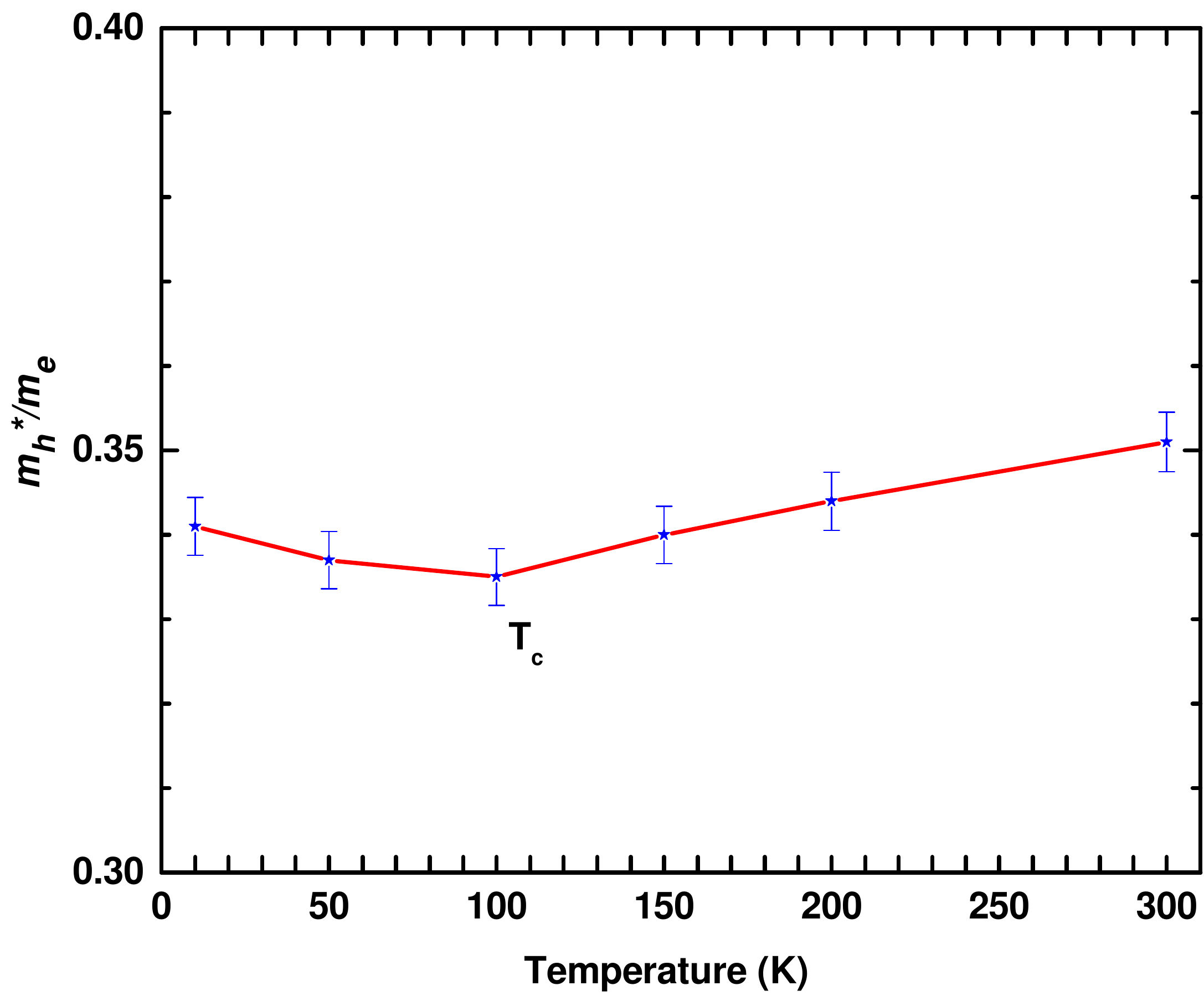}
\caption{\label{fig:Effectivemass} (Color online) Temperature-dependent hole effective mass of Pb${}_{0.77}$Sn${}_{0.23}$Se.}
\end{figure}

We note that on account of the doping, the Fermi level is far from the putative Dirac point, and, moreover that the optical conductivity
spectrum from which this effective mass is derived is an angular average over the Fermi surface and an energy average over a scale set by
the relaxation rate, $1/\tau$.

\subsection{Energy-loss function}

Finally, the temperature dependence of the longitudinal response of this material, namely the loss function $L(\omega)$, is plotted against
 frequency in Fig. 7. The free-carrier peak in the loss function is a good estimate of the screened Drude plasma frequency. This estimate is
  in good agreement with the zero crossing of $\varepsilon_1$ in Fig. 4 as well as the value calculated by fitting with the Drude-Lorentz model,
  discussed in the next section.

\begin{figure}[H]
\centering
\includegraphics[width=3.5 in,height=3.5 in,keepaspectratio]{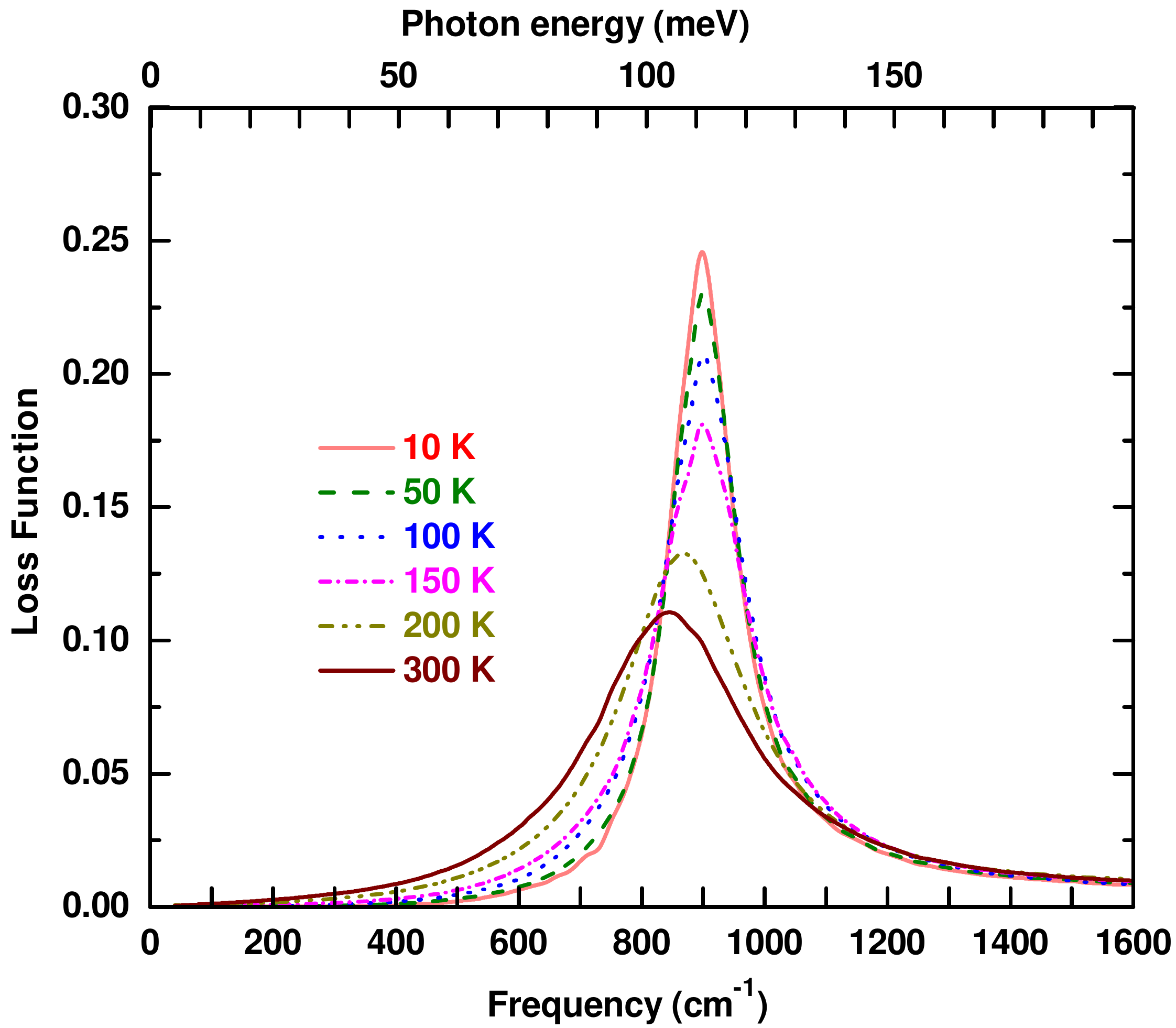}
\caption{\label{fig:lossfunction} (Color online) Temperature-dependent energy loss function of Pb${}_{0.77}$Sn${}_{0.23}$Se.}
\end{figure}

\subsection{Drude-Lorentz model fits}

We used the Drude-Lorentz model to fit the reflectance data. The Drude component characterizes the free carriers and their dynamics at zero frequency
 whereas the Lorentz contributions are used for the optically-active phonon in the far-infrared region along with valence intraband and interband
 transitions in the mid and near infrared. The dielectric function is:

\begin{equation}
\varepsilon (\omega)=\varepsilon_{\infty}- \frac{\omega _{p}^{2} }{\omega^{2}+i\omega /\tau } +
\sum _{j=1}^6\frac{\omega_{pj}^{2} }{\omega_{j}^{2}-\omega^{2} -i\omega \gamma_{j} }
\end{equation}

where the first term represents the core electron contribution (transitions above the measured range, the second term is free carrier contribution
characterized by Drude plasma frequency $\omega _{p}$ and free carrier relaxation time $\tau $ and the third term is the sum of six Lorentzian
oscillators representing phonons, valence intraband, and interband electronic contributions. The Lorentzian parameters are the $j$th oscillator
plasma frequency $\omega _{pj}$, its central frequency $\omega _{j}$, and its linewidth $\gamma _{j}$. This dielectric function model is used in
a least-squares fit to the reflectance.
\begin{figure}[H]
\centering
\includegraphics[width=3.4 in,height=3.4 in,keepaspectratio]{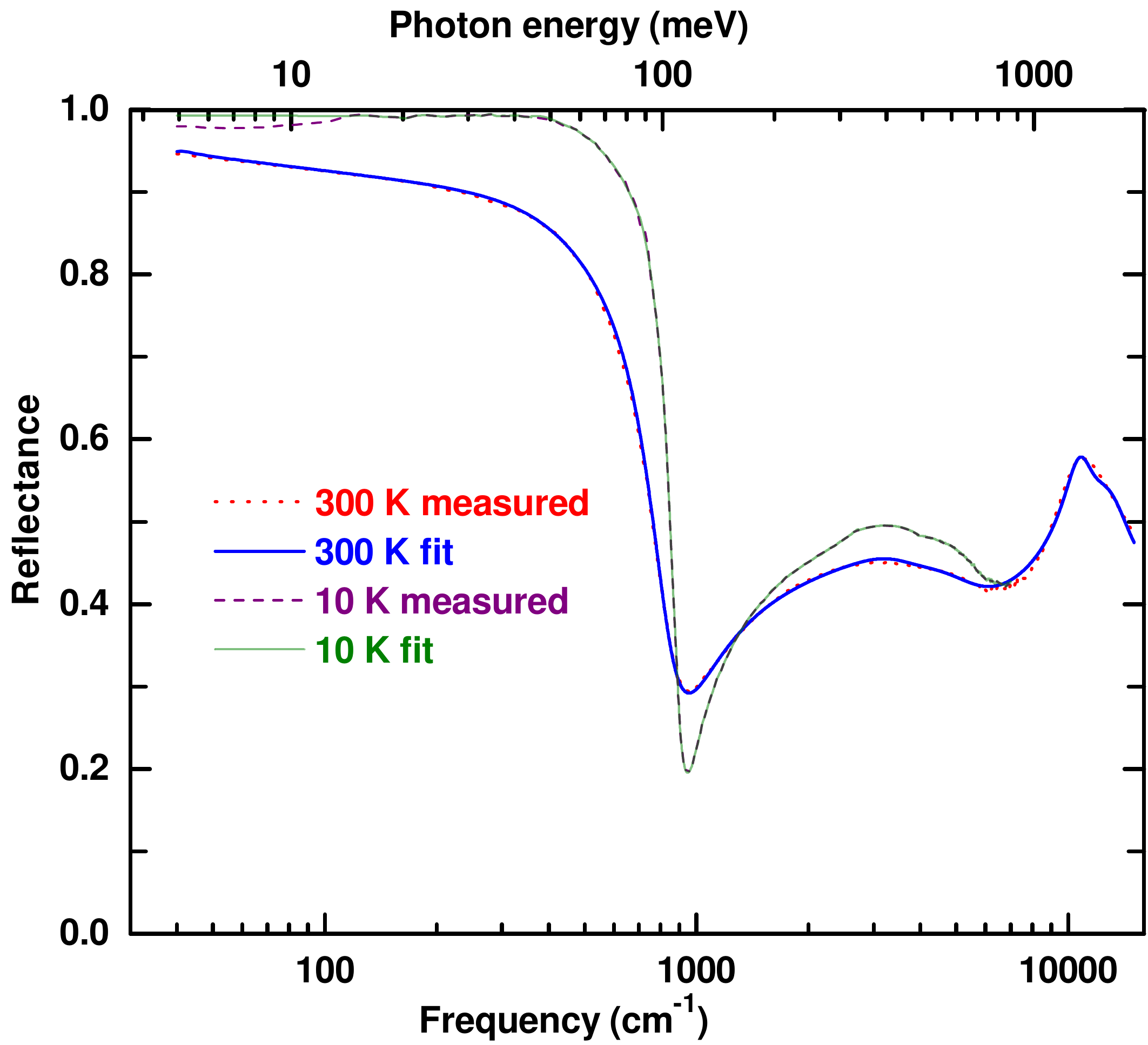}
\caption{\label{fig:reflfit} (Color online) Drude-Lorentz fit to the reflectance of Pb${}_{0.77}$Sn${}_{0.23}$Se at 300~K and 10~K.}
\end{figure}

\begin{figure}[H]
\centering
\includegraphics[width=3.4 in,height=3.4 in,keepaspectratio]{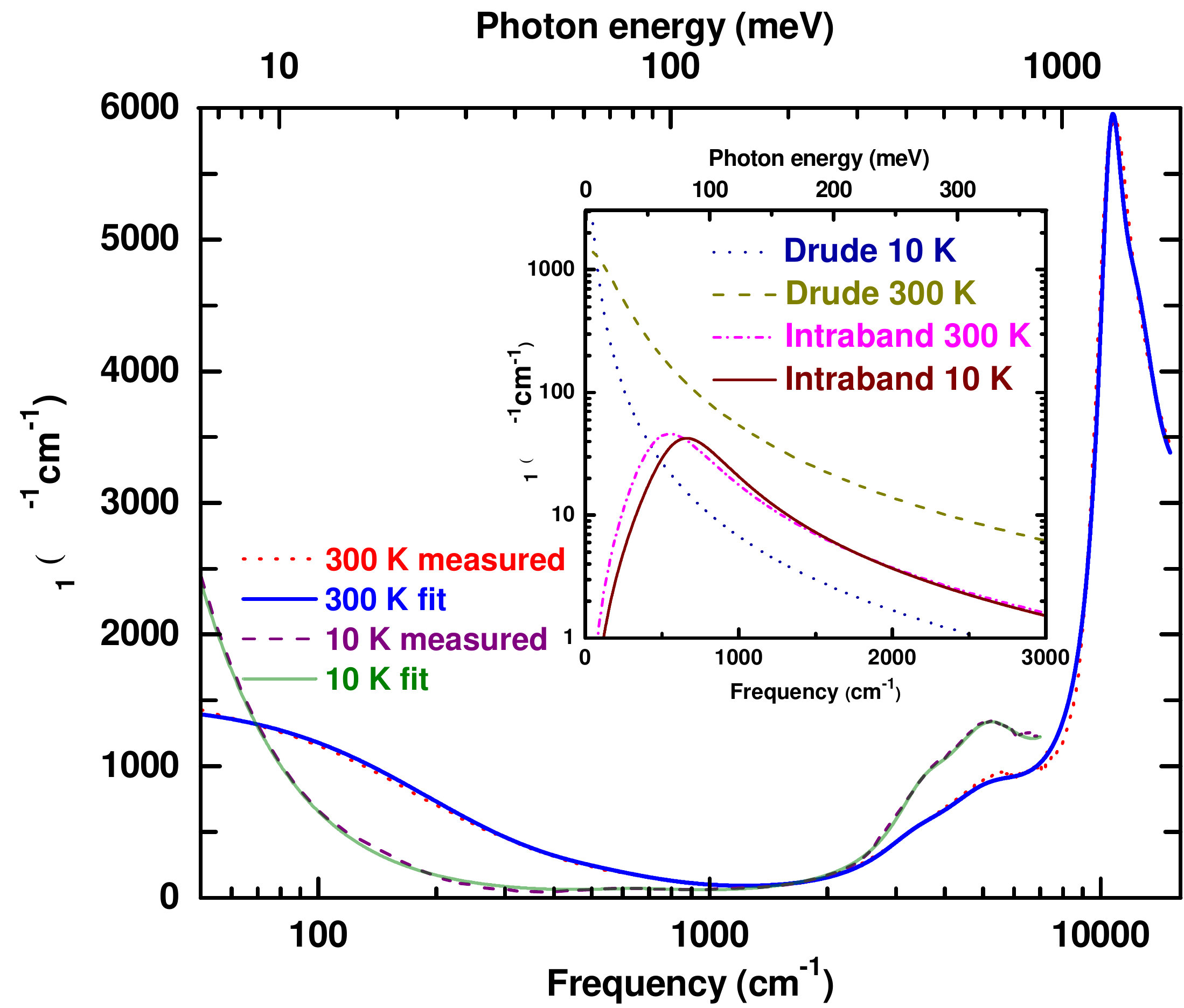}
\caption{\label{fig:conductivityfit} (Color online) A comparison of the Kramers-Kronig-derived optical conductivity and the conductivity calculated
 from the Drude-Lorentz fit to the reflectance of Pb${}_{0.77}$Sn${}_{0.23}$Se. Results are shown for 10 and 300~K. The inset shows the Drude and
  the valence intraband contributions at the two temperatures.}
\end{figure}

Figure 8 compares the calculated   and measured reflectance at 10~K and 300~K;  the 300~K fit is shown up to 15,500 cm${}^{-1}$ whereas the 10~K
data are fitted to 7000 cm${}^{-1}$. Similar quality fits were obtained at all other temperatures.

The comparison between the Kramers-Kronig-derived and the Drude-Lorentz-model conductivity spectra at 10~K and 300~K is shown in Fig. 9.
The parameters are the ones used to fit the reflectance. That the agreement is good gives us confidence in both procedures. The inset graph
shows the Drude contribution and the valence intraband contributions  at 10~K and 300~K. The valence intraband peak is small and dispersive
but is necessary for a good description of the data. Table I lists the 21 parameters used during the fitting routine to fit the reflectance
and the conductivity spectra at 300~K.

\begin{table}[h]
\caption { Drude-Lorentz parameters for Pb${}_{0.77}$Sn${}_{0.23}$Se at room temperature (300~K). }
\centering
\begin{tabular}{c c c c}
\hline\hline
Modes assignment & Oscillator  & Central  & Linewidth \\
in fitting & strength $\omega _{p}$ & frequency $\omega _{j }$  & $\gamma _{ j}$   \\
routine  &  (cm${}^{-1}$) & (cm${}^{-1}$) &  (cm${}^{-1}$) \\[1ex]
\hline
Drude & 4160 & 0 & 195  \\
Optical phonon & 240 & 40 & 5.5 \\
Valence intraband & 1230 & 550 & 560 \\
Interband 1 & 4550 & 3500 & 1900 \\
Interband 2 & 11,000 & 5120 & 3950 \\
Interband 3 & 20,630 & 10,620 & 1950 \\
Interband 4& 23,750 & 12,350 & 4150 \\ [1ex]
\hline\hline
\end{tabular}
\end{table}

\begin{figure}[H]
\centering
\includegraphics[width=3.2 in,height=3.2 in,keepaspectratio]{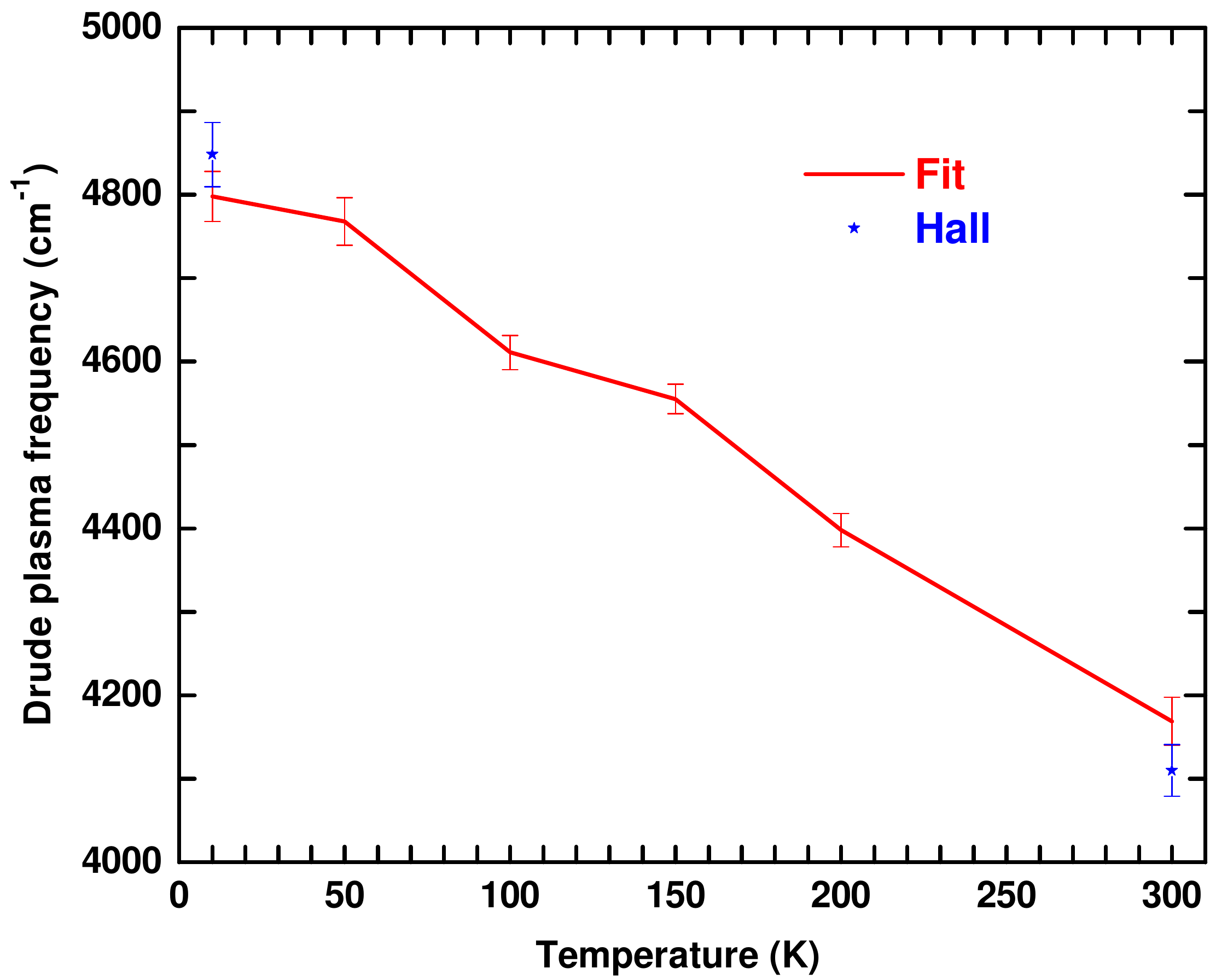}
\caption{\label{fig:drudeplasmafrequency2} (Color online) Comparison between the Drude-Lorentz model fitted and Hall-derived Drude plasma frequency (10~K-300~K)}
\end{figure}

Figures 10--13 show the temperature dependence of selected parameters from the fits. First, Fig. 10 shows the Drude plasma frequency,
whose strength decreases as temperature increases. We also show a plasma frequency calculated from the   Hall measurement
  $\omega _{p} = \sqrt{4 \pi n e^2/m_h^*}$ where $n$ comes from the Hall data and $m_h^*$ from the sum rule analysis for the free carriers.
  Within the error bars, the two results agree.

\begin{figure}[H]
\centering
\includegraphics[width=3.6 in,height=3.6 in,keepaspectratio]{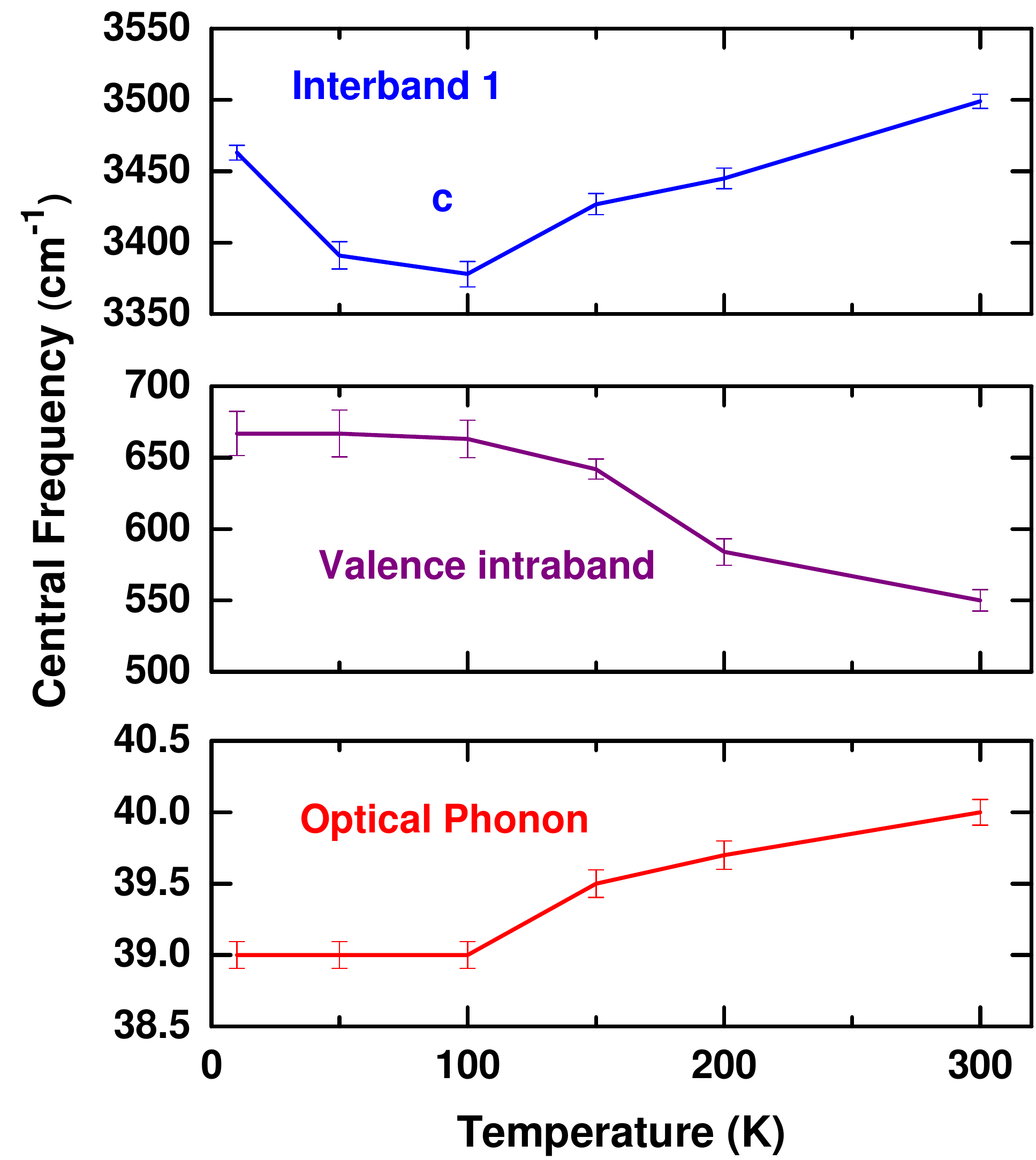}
\caption{\label{fig:centralfrequency} (Color online) Temperature dependence of the central frequency for the fitted Lorentz oscillators.}
\end{figure}

Secondly, the temperature dependence of the central frequencies of the lowest interband transition, the valence intraband transition and the
optical phonon is shown in Fig. 11. The lowest interband transition has its minimum value at $100\pm 25$~K. As temperature increases,
the valence intraband oscillator shifts towards lower frequencies whereas the phonon mode in the far-infrared range shifts slightly towards
higher frequencies.

The linewidth of many of these modes suggests significant overlap amongst them, especially in the midinfrared region.
Figure 12 shows the temperature dependence of the linewidth $\gamma$ of these transitions. The Drude relaxation rate (bottom panel) becomes as
small as 17 cm${}^{-1}$ (well below our experimental low frequency limit) at 10~K,  increasing with temperature to 195 cm${}^{-1}$ at 300~K.
The corresponding hole mobility $\mu = e\tau/m_h^*$ decreases from 1630 cm${}^2$/Vs at 10 K to 140 cm${}^2$/Vs at 300 K. Based on these hole mobility
 values, the optical conductivity $\sigma = n e \mu$ is estimated to be 23,200 $\Omega$${}^{-1}$cm${}^{-1}$ at 10~K, decreasing to 1320 $\Omega$$^{-1}$cm${}^{-1}$ at 300~K.
  These estimates are in good agreement with the conductivity values found from the fitting routine: 21,800 $\Omega$${}^{-1}$cm${}^{-1}$
   at 10~K, decreasing to 1490 $\Omega$${}^{-1}$cm${}^{-1}$ at 300~K. The linewidth of every other oscillator shows a similar temperature trend
    except for the valence intraband mode which has a weak (and not very significant) minimum at 100 K.

\begin{figure}[H]
\centering
\includegraphics[width=3.9 in,height=3.9 in,keepaspectratio]{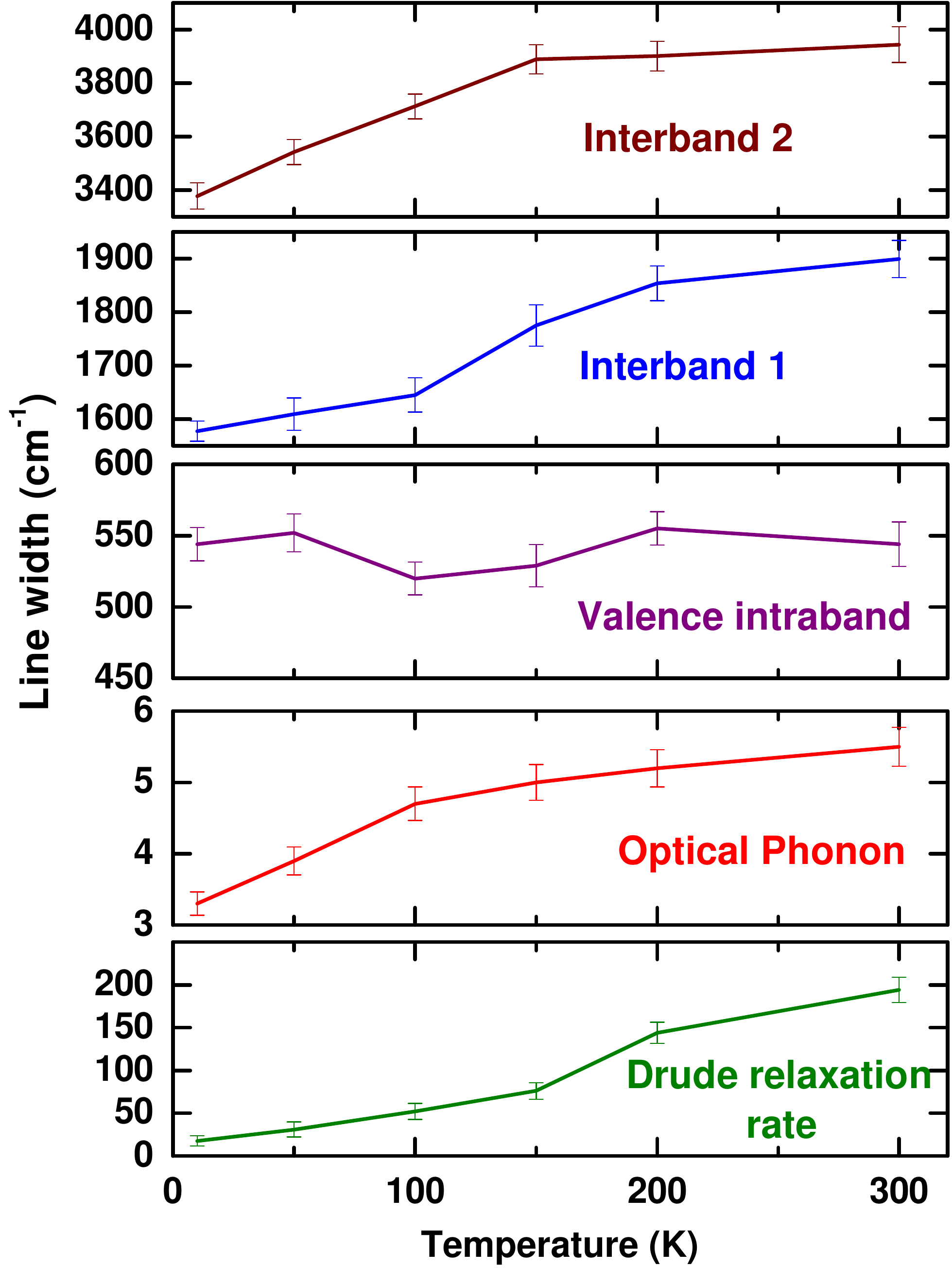}
\caption{\label{fig:linewidth} (Color online) Temperature dependence of the linewidth for the fitted Lorentz oscillators.}
\end{figure}

\begin{figure}[H]
\centering
\includegraphics[width=3.8 in,height=3.8 in,keepaspectratio]{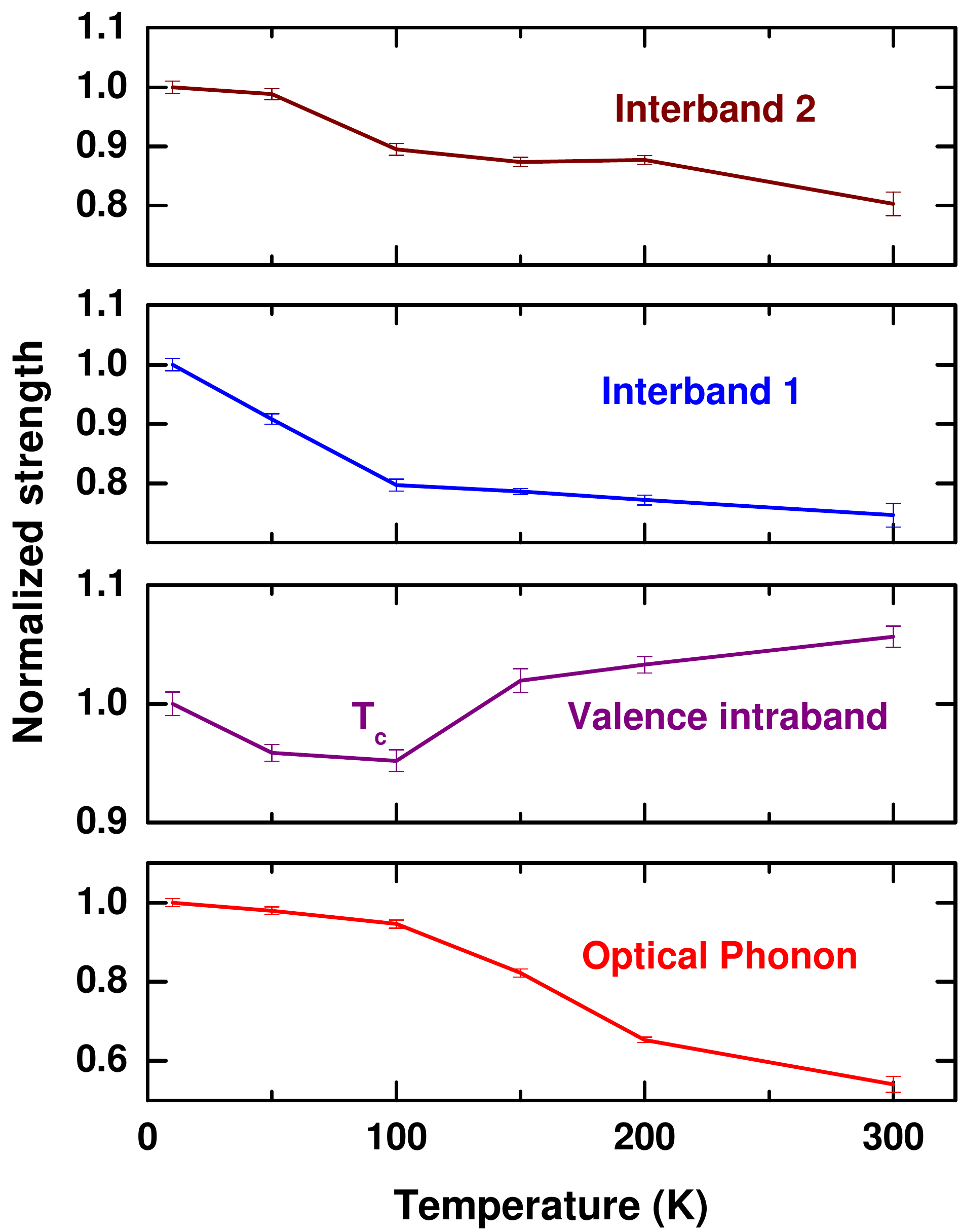}
\caption{\label{fig:Normalizedstrength} (Color online) Temperature dependence of the oscillator strength (normalized by the 10~K value) for the
 fitted Lorentz oscillators.}
\end{figure}

Finally, the temperature dependence of the oscillator strength for several oscillators is shown in Fig. 13. The strength has been normalized
by its value at 10~K. (The Drude oscillator strength is shown in Fig. 10.) Three  oscillators have their highest strength at the lowest
temperature; it  decreases monotonically as temperature increases, with a change of slope around 100~K. The valence intraband mode is different:
its  shows a clear  minimum at 100~K and is higher at 300~K than at 10~K.

\section{DISCUSSION}

\subsection{ Free carrier characteristics and band inversion }

 A scattering rate calculation for a p-type Si${}_{1-x}$Ge${}_{x}$ alloy predicts quite a high scattering rate.\cite{Joyce}
 Impurity scattering is expected to be strong on account of the disorder associated with the alloy. In contrast, our Pb${}_{0.77}$Sn${}_{0.23}$Se
 single crystal has a  quite small free carrier scattering rate at low temperature. Indeed, the free-carrier scattering rate (shown in the bottom
  panel of Fig. 12) remains temperature dependent below 50 K.  Considering the alloy form of the material with large stoichiometric imbalance,
  this result is rather surprising.

Previous studies have shown that the extrinsic carrier density in Pb${}_{1-x}$Sn${}_{x}$Se affects not only the Hall coefficient but also
its temperature dependence. For lightly doped systems, $R_{H}$ changes sign from positive to negative as temperature is increased, mainly
 due to the increasing $n$ type contribution from thermally excited carriers. On the other hand, for heavily doped systems, the Hall coefficient
  is positive throughout and has weak temperature dependence.\cite{Dixon} As temperature is increased, due to increasing thermal energy,
  the number of free carriers increases. However an increase in the band gap due to increasing temperature also reduces the number of thermally
  excited carriers. The value of the Hall coefficient depends on the relative dynamics of these two opposing mechanisms. The slight increase in
  the Hall coefficient with temperature in our sample could be due to an imbalance between these two competing processes.

Temperature dependent ARPES measurements of the (100) surface of Pb${}_{0.77}$Sn${}_{0.23}$Se monocrystals have been interpreted as finding
 band inversion in the material around 100~K.\cite{Dziawa} The band curvature at the L point changes during band inversion and band dispersion
 becomes almost linear during gap closure at 100~K. The temperature dependence of the carrier effective mass (shown in Fig. 6) does show a weak minimum,
  in agreement with this idea.

\subsection{Minimum band gap and the valence intraband transition}

The phenomenon of first shrinking and then expansion of band gap as temperature is decreased should affect the minimum absorption edge
directly. Fig. 11 shows that the central frequency of the lowest interband transition has a minimum at 100~K. In addition, the onset of
this interband transition, estimated by linear extrapolation of the low-energy edge in the Kramers-Kronig-derived optical conductivity to
 zero conductivity suggests a similar trend. An ARPES study also finds a minimum band gap at 100~K.\cite{Dziawa}

We can point to other features in the optical spectrum that present themselves around 100~K: the frequency of the optically-active phonon (Fig. 11)
 and the linewidth and the spectral weight of the valence intraband transition (Figs. 12 and 13).

\subsection{Electronic structure calculations}

The energy band structure of Pb${}_{0.75}$Sn${}_{0.25}$Se was calculated within density-functional theory using the VASP
 package.\cite{Kresse93,Kresse94,Kresse96,Kresse1996} The input structure consists of 8 atoms cell with 3 Pb, 1 Sn and 4 Se atoms.
 The calculation was based on a generalized gradient approximation (GGA) for the exchange-correlation potential using the Perdew, Burke, and
 Ernzerhof formalism\cite{John} along with spin-orbit coupling for both geometric and electronic optimization. We used an energy cutoff of 500 eV,
  10${}^{-6}$~eV energy minimization, and $8\times8\times8$ $k$-points. The calculation gave a lattice constant of 6.16~{\AA}, an overestimate by
   0.09 ~{\AA} compared with the measured lattice constant.\cite{Haas} The electronic band structure was calculated along high-symmetry $k$-points in
    the Brillouin zone with the Fermi energy fixed to zero ({\it i.e.,} an undoped crystal). As shown in Fig. 14, the band structure calculation indicates
    that the material has a direct band gap at the L-point in the Brillouin zone with an energy gap of 0.20 eV. Fig. 15 shows an enlarged view of the
    conduction band minimum and the valence band maximum  at the L point.

\begin{figure}[H]
\centering
\includegraphics[width=3.5 in,height=3.5 in,keepaspectratio]{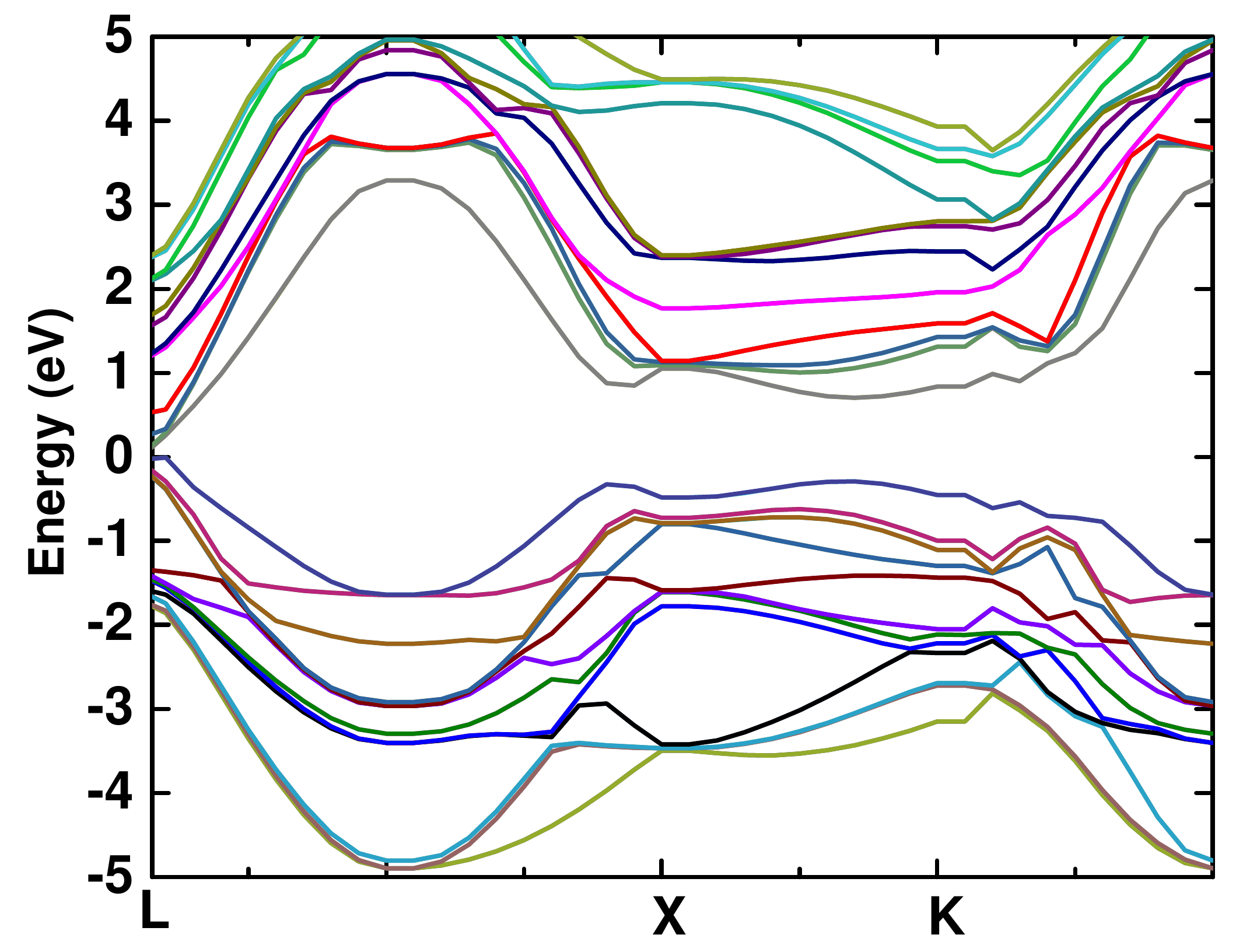}
\caption{\label{fig:bandstructure} (Color online) Calculated electronic band structure of Pb${}_{0.75}$Sn${}_{0.25}$Se.}
\end{figure}

The spin-orbit interaction is important in determining the actual band ordering, band splitting and the details of the electronic structure
in class IV-VI narrow-gap semiconductors. Fig. 15 shows the splitting of the valence band into 3 valence sub-bands. The splitted valence bands
are named as split-off, heavy-hole and light-hole bands.\cite{Jacques, Braunstein}

\begin{figure}[H]
\centering
\includegraphics[width=3.2 in,height=3.2 in,keepaspectratio]{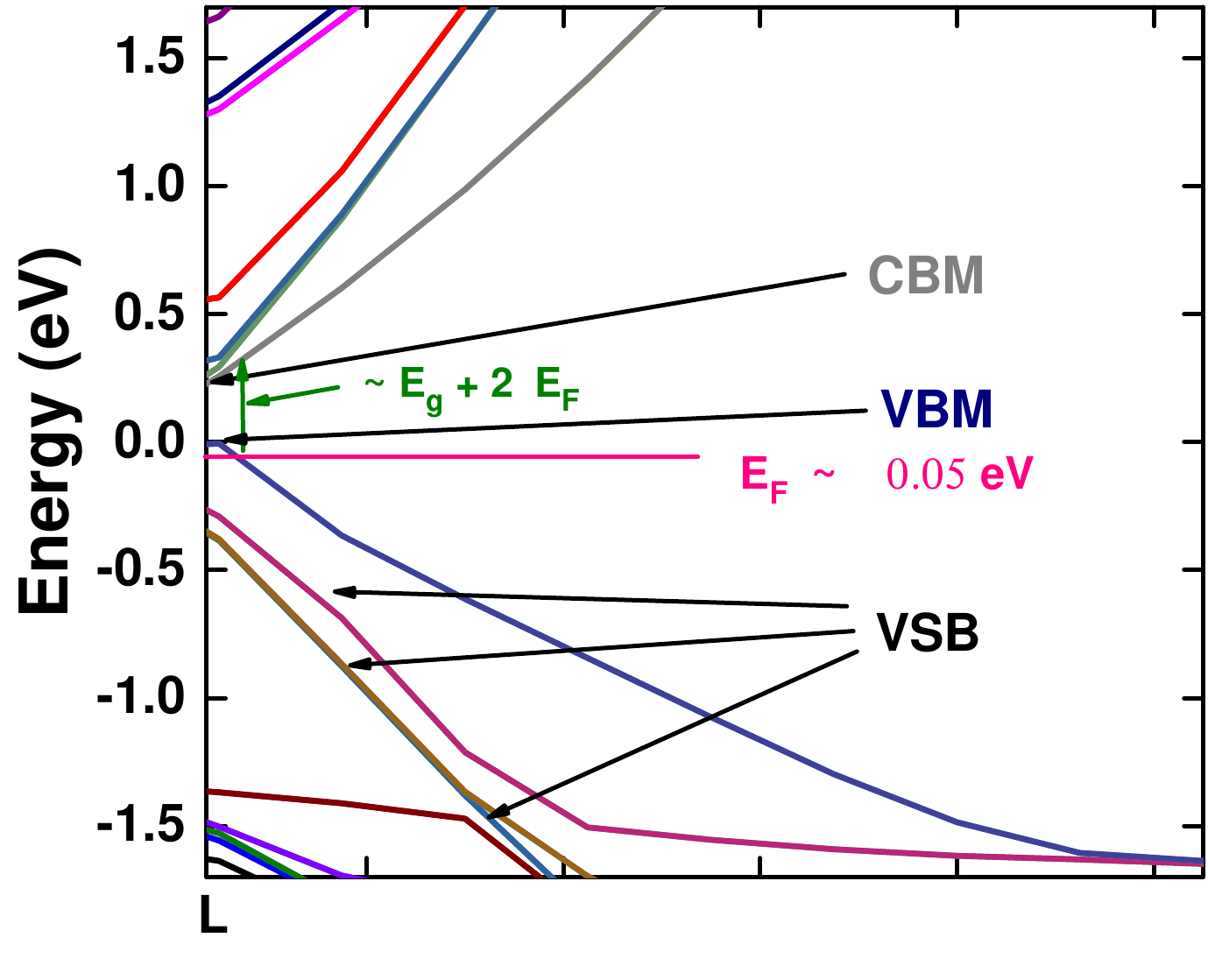}
\caption{\label{fig:bandstructure2} (Color online) Enlarged view at the L point of the low-energy part of the band structure showing the
splitting of the valence band. CBM = conduction band minimum; VBM = valence band maximum; VSB= valence sub-band.}
\end{figure}

This VASP calculation also produced the density of states. Using the carrier density from experiment and this density of states,
we can estimate that the Fermi energy in our doped sample lies around $E_F = -50$~meV.  The lowest interband transition takes place
from $E_Fi$ to the empty state in the conduction band at the same $k$-point . The band structure calculation then predicts this absorption
 to start around 0.30 eV (2420 cm${}^{-1}$). This prediction is in good agreement with the estimate of 2300 cm$^{-1}$ from the threshold for
  the interband transitions seen in the optical conductivity.

In addition to the optical interband transitions between the valence and the conduction band, there are also optical transitions within the
valence band, the valence intraband transition discussed above. These transitions are seen in a number of p-type semiconductors, when the top of
 the valence band is populated with holes. It is then possible to make intraband transitions from low-lying sub-bands to states above the Fermi level.
 These transitions have been observed in a number of semiconductors, including Ge, GaAs, and InSb.\cite{Kaiser, Gobeli, Grundmann} The single crystal
  of Pb${}_{0.77}$Sn${}_{0.23}$Se shows similar valence intraband transitions whose temperature dependence is shown in Figs. 11--13.

\section{CONCLUSIONS}

Hall measurements of Pb${}_{0.77}$Sn${}_{0.23}$Se single crystals disclose the p-type nature of the material and show that the carrier density
 decreases from $9\times10^{19}$~cm${}^{-3}$ at 10~K to $6\times10^{19}$~cm${}^{-3}$ at 300~K. Temperature-dependent optical reflectance measurement
 shows degenerate semiconductor behavior. At 10~K, the screened Drude plasma minimum lies around 930 cm${}^{-1 }$; it shifts slightly towards lower
  frequency as temperature increases.

In the midinfrared and near-infrared region, the spectra show valence intraband transition and valence to conduction band excitations.
The temperature dependence of the lowest-energy interband transition energy shows a minimum  at 100~K. This behavior suggest band inversion in the material,
 also predicted by many other previous studies. The  effective hole  mass  also has  a minimum value at 100~K consistent with linear band dispersion at the
 L point in the Brillouin Zone. Theoretical calculations provide the details of electronic band structure, supporting the existence of valence intraband
 transition. Our experimental and theoretical studies suggest that material undergoes a temperature-driven band inversion at the L point as also predicted
 by previous study.\cite{Dziawa}

\begin{acknowledgments}
The author wish to thank Xiaoxiang Xi for his constructive discussions on material's band structure. This work was supported by the DOE through contract No.
DE-AC02-98CH10886 at Brookhaven National Lab and by the NSF DMR 1305783 (AFH).
\end{acknowledgments}

\bibliography{PbSnSe}
\end{document}